\documentclass[12pt,preprint]{aastex}

\newcommand{\be}{\begin{eqnarray}}
\newcommand{\ee}{\end{eqnarray}}
\newcommand{\etal}{{\it{et al.}}}
\newcommand{\smass}{M_{\odot}}

\shorttitle{To circularize or not to circularize?}
\shortauthors{Hashimoto, Funato and Makino}

\begin{document}

%
%
\title{
To circularize or
not to  circularize? ---
Orbital evolution of satellite galaxies
}

%
%

\author{Yoshikazu Hashimoto}
\affil{Department of Astronomy,
Graduate School of Science, the University of Tokyo,
    Tokyo, 113, Japan}
\email{hasimoto@astron.s.u-tokyo.ac.jp}
\author{Yoko Funato}
\affil{
General Systems Studies, Graduate Division of International and
Interdisciplinary Studies, University of Tokyo, Tokyo, 153, Japan}
\email{funato@chianti.c.u-tokyo.ac.jp}
\and
\author{Junichiro Makino}
\affil{Department of Astronomy,
Graduate School of Science, the University of Tokyo,
    Tokyo, 113, Japan}
\email{makino@astron.s.u-tokyo.ac.jp}

%
%

\begin{abstract}

We investigated the orbital evolution of satellite galaxies using
numerical simulations.  It has been long believed that the orbit
suffers circularization due to the dynamical friction from the
galactic halo during orbital decay. This circularization was confirmed
by numerous simulations where dynamical friction is added as external
force. However, some of the resent $N$-body simulations demonstrated
that circularization is much slower than expected from approximate
calculations. We found that the dominant reason for this discrepancy is
the assumption that Coulomb logarithm $\log \Lambda$ is constant,
which has been used in practically all recent calculations. Since the
size of the satellite is relatively large, accurate determination of
the outer cutoff radius is crucial to obtain good estimate for the
dynamical friction. An excellent agreement between $N$-body
simulations and approximate calculations was observed when the outer
cutoff radius is taken to be the distance of the satellite to the
center of the galaxy. When satellite is at the perigalacticon, the
distance to the center is smaller and therefore $\log \Lambda$ becomes
smaller. As a result, the dynamical friction becomes less effective.

We apply our result to the Large Magellanic Cloud.  
We found that the expected lifetime of the LMC is twice as
long as that would be predicted with previous calculations.  Previous
study predicts that the LMC will merge into the Milky Way after 7 G
years, while we found that the merging will take place after 14 G
years from now.  Our result suggests that generally satellites formed
around a galaxy have longer lifetime than previous estimates.

\end{abstract}


%
%


\section{Introduction}

Recent observations have revealed that there are many satellite
galaxies around the Milky Way. In the hierarchical clustering
scenario, it is expected many of such dwarf satellites are formed. In
fact, one of the most serious problems with the present hierarchical
clustering scenario is that it predict too many satellite galaxies,
about a factor of 10 more than the number observed in the Local group
({\it e.g.}, Moore \etal, 1999). A number of explanations,
including exotic theories which relies on hot or self-interacting dark
matter, have been proposed.

In this paper, we go back to the basic problem: how long are the
satellites lives ?  In other words, how do the orbits of satellites
evolve through interaction with the gravitational field of its parent
galaxy? The dominant driving force of the evolution is the dynamical
friction. For satellites like the LMC-SMC pair and the Sagittarius
dwarf, there are many detailed studies of their orbital evolution, in
which the dynamical friction is included as the external force
operating on the center-of-mass motion of the satellite.  Well known
works include Murai and Fujimoto (MCs, 1980) and Ibata and Lewis
(Sagittarius, 1998). In both of these studies, and in all other
studies where the dynamical friction formula is used, significant
circularization of the orbit of the satellite was observed. This
circularization is the natural result of the fact that the dynamical
friction is proportional to the local density of the background stars,
and therefore the strongest at the perigalacticon.

However, recent $N$-body simulations of the orbital evolution of
satellites resulted in rather counter-intuitive result. Van den Bosch
et al (1999, hereafter BLLS) performed the $N$-body simulation of the
satellite, where the parent galaxy was modeled directly as
self-consistent $N$-body system. The satellite is modeled as one
massive particle with spline potential softening used in PKDGRAV
(Dikaiakos \& Stadel, 1996).  They investigated the
evolution of the orbit for wide variety of model parameters such as
the mass of the satellite and initial orbital eccentricity. They
observed practically no circularization in any of their simulations.

Jiang and Binney (2000, hereafter JB) performed fully self-consistent
simulation of the satellite, where both the parent galaxy and the
satellite are modeled as self-consistent $N$-body systems. They
compared their result with the result of approximate model in which
the usual dynamical friction formula is used. Though they argued that
the agreement is good, from their figure 3 it is clear that
approximate models suffer stronger circularization and evolve faster
than their $N$-body counterpart.

Neither of above two papers discussed the reason of this rather
serious discrepancy between the result of $N$-body simulations and
previous analytic prediction.  The purpose of this paper is to
understand its cause. In section 2, we describe our model experiment
designed to reproduce the discrepancy observed by BLLS and JB. In
section 3 we show our result. Our result is consistent with both of
the previous works. $N$-body simulation showed only marginal
circularization but approximate calculation using dynamical friction
formula showed strong circularization. In section 4, we investigate
the reason. There are several possible candidates for the reason. We
consider a few of them, and found that a simple modification of the
conventional form of the dynamical friction formula results in a quite
remarkable improvement of the agreement between $N$-body and
approximate calculations. In section 5 we apply our formalism to the
LMC. We found that the orbital evolution becomes
significantly slower than prediction by previous calculations using
conventional formula. For example, the lifetime of LMC was 7 Gyr with
conventional formula, but is 14 Gyr with our formalism. We also
discuss the implication of our result to the so-called ``dwarf
problem''.

\section{Numerical Simulation}

We carried out a set of numerical simulations to see whether the results
obtained by BLLS and JB are really true or not. In this section, we
describe the models we used.

\subsection{$N$-body simulation}

We performed $N$-body simulations of the evolution of a satellite
orbiting in a massive dark halo of a galaxy.

The massive halo is composed of $N$ equal mass particles, while the
satellite dwarf is modeled by a single particle with a certain
softening length. The softening is used to mimic the finite size of  the
satellite.

We adopted a King model of the concentration ratio $\Psi_0 = 9$ as a
model of the galactic halo. The system of units is the Heggie unit
(\citet{hm86}) where the gravitational constant $G$ is $1$, the mass
and the binding energy are $1$ and $0.25$, respectively.

JB used a composite disk+halo model in which the halo is
expressed by particles and the disk is assumed to be rigid. BLLS used
a  single spherical halo. In both works, the halo density profile has
the form
\begin{equation}
\rho = \rho_0 \frac{r_c^2\exp[-(r/r_t)^k]}{r_c^2+r^2},
\label{eq:modelden}
\end{equation}
where $r_c$ and $r_t$ are the core radius and the outer scale radius
of the halo and $\rho_0$ is the central density of the halo. BLLS
adopted $k=2$ while JB adopted $k=1$.

We did not follow the models.  The standard dynamical friction formula
is derived for the case of field stars with the Maxwell distribution.
However, the distribution function associated with
eq. (\ref{eq:modelden}) is rather different from the Maxwell
distribution.
This may cause difference in the effect of the dynamical
friction. Also, the distribution function would relax to the
Maxwellian through two-body relaxation, causing a small change in both
the distribution function and the density profile.

In addition, the range of radius for which the density slope is
approximately $-2$ is rather narrow with this model, since the slope
is noticeably shallower than $-2$ for $r \le 10 r_c$.

The distribution function of the King model is a simple lowered
Maxwellian. Therefore the agreement with the true Maxwellian is very
good within the half-mass radius. Also, since the distribution
function is practically as close as the true Maxwellian as we can
make, thermal relaxation is minimized, though it still presents (see
e.g., Quinlan 1996). Also, the King model with $\Psi_0=9$ has fairly
wide range of radius in which the slope of the density is
approximately $-2$. So it is a fairly good model for a spherical halo
with flat rotation.

The satellite galaxy is modeled by a single particle with mass
$M_{{\rm s}}$ and softening length $\epsilon_{{\rm s}} $.  The force
on the satellite from a particle in the halo is calculated as follows
\be
{\mathbf F} & = & -\frac{G m M_{{\mathrm s}}
({\mathbf r}_{{\mathrm {sat}}} -{\mathbf r}_{{\mathrm {halo}}})}
{(|{\mathbf r}_{{\mathrm {sat}}} -{\mathbf r}_{{\mathrm {halo}}}|^2 +
\epsilon_{{\mathrm s}}^2
+ \epsilon_{{\mathrm {halo}}}^2)^{3/2}}.
\ee
Here $\epsilon_{{\mathrm {halo}}}$ is the softening length for the particle
in the galactic halo. The value of the gravitational constant $G$ is
$1$ in the standard units.

In table \ref{tbl-1} we summarized the model parameters and initial
conditions of our $N$-body simulations.

In our simulations, equations of motions of all particles in a dark
halo and the satellite, {\it i.e.}, $N + 1$ particles, are integrated
self-consistently. In other words, the dynamical friction effect from
halo particles to the satellite is included naturally.

The number of halo particles $N$ used in the simulations shown in this
paper is 32768. We varied $N$ from $8192$ to $65536$, and found any
noticeable difference in the orbit of the satellite.

We used GRAPE6 to calculate the acceleration. We adopted simple
$O(N^2)$ direct summation, to avoid any possible numerical artifact
caused by the approximations made in force calculation. BLLS used the
treecode and JB used a composite grid-based code. We do not think the
numerical method caused the difference, but we want to be absolutely
sure that our $N$-body simulation is as accurate as possible.

We integrated the orbits of the satellites and halo particles using
the standard leapfrog scheme with a shared stepsize, $\Delta t =
0.03125$.  The error in total energy within $10^{-3}$, which is small
enough to observe the orbit evolution of satellites (Hashimoto \etal,
2002).

\subsection{Semi-analytic Integration}

We performed semi-analytic
calculations to follow the evolution of satellite orbits.

In these calculations, the model of the satellite is the same as in the
$N$-body simulations, {\it i.e.}, a single particle with mass $M_{{\rm
s}}$ and the softening length $\epsilon_{{\rm s}}$.

Instead of being represented by $N$ particles, the potential of the
galactic halo is evaluated by using the gravitational potential of
King 9 model with the same mass and scales as those adopted in
$N$-body simulations.

In this integration, the force to the satellite due to the dynamical
friction from the halo is evaluated by using an analytical formula,
too.

For the dynamical friction formula, we follow JB (and also Murai and
Fujimoto) to use the standard  ``Chandrasekhar's
dynamical friction formula''. It is expressed as 
\be
\frac{d{\mathbf v}}{dt
}= -16\pi^2 G^2 m(M_{{\mathrm s}}+ m) \ln \Lambda 
\frac{\int_0^{v_{{\mathrm {max}}}} f(v) d{\mathbf v}}
{|{\mathbf v}|^3} {\mathbf v},
\label{eqn:dff1}
\ee
where $m$ and $M_{{\rm s}}$ are the masses of a component of host galaxy 
and its satellite galaxy \citep{Ch43,BT87}.
Here  $\ln \Lambda $ is the Coulomb logarithm 
\begin{equation}
\ln \Lambda = \ln ({R_{\mathrm {halo}}}/
{\epsilon_{\mathrm s}{V_{\mathrm s}}^2}),
\label{eqn:clambda}
\end{equation}
where $R_{{\mathrm {halo}}}$ is the scale length of the galactic halo. 
This formula has been adopted by many semi-analytic studies of the
orbital evolution of satellite galaxies ({\it e.g.}, Murai and
Fujimoto 1980; Helmi and White, 1999; Johnston \etal, 1995). It is
also used in cosmological studies of galaxy formation in order to
estimate the merging time scale of satellite galaxies ({\it e.g.},
Kauffmann, \etal, 1994).

\section{Result}

\begin{figure}
\plotone{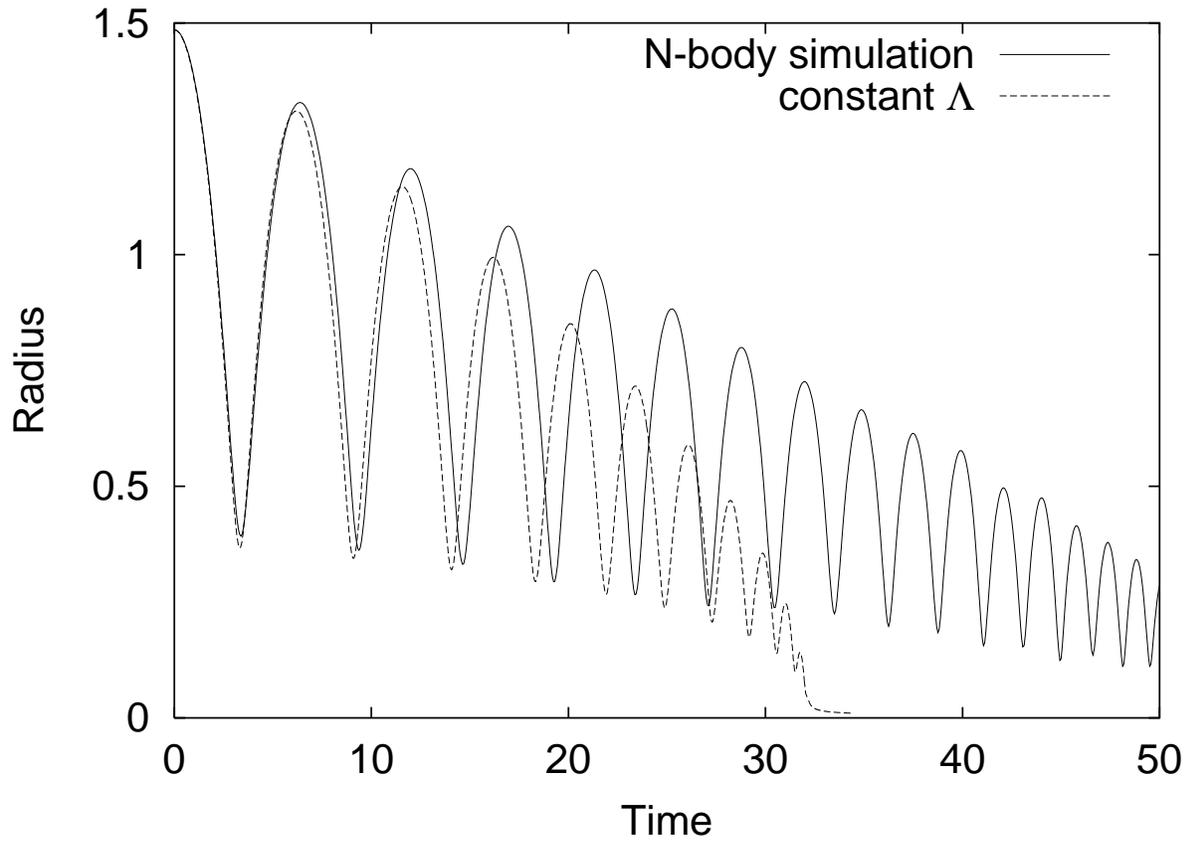}
\caption{Time evolution of radius of satellite position from the
galaxy center.  Solid: result of $N$-body simulation. Dashed:
semi-analytical integration using constant $\Lambda$.}
\label{fig:1}
\end{figure}

Figure \ref{fig:1} shows the orbital evolution of a model satellite
galaxy.  The ordinate and abscissa are the distance of the satellite
from the center of the galaxy and time in the $N$-body units.  The
solid and dashed curves correspond to the result of $N$-body
simulation and that of semi-analytic model with standard dynamical
friction formula (\ref{eqn:dff1}).

In Figure \ref{fig:1} two curves are in good agreement only for a
first few dynamical times. After a few orbits, two curves deviate  from 
each other. 
Figure \ref{fig:1} shows that the orbital decay calculated with formula
(\ref{eqn:dff1}) is faster than that obtained by $N$-body
simulation. If one measure the orbital eccentricity, it is clear that 
$N$-body result shows only a small change in the eccentricity, while
semi-analytic result shows significant circularization.

Thus, even though we used completely different initial models and
numerical method, we confirmed previous results by BLLS and JB that
$N$-body simulation shows little circularization while semi-analytical 
calculation with standard dynamical friction formula shows strong
circularization. In the next section, we discuss the possible causes
of this discrepancy. 

\section{Possible causes of discrepancy}

Since we have obtained quite different results with $N$-body and
semi-analytic models, {\it at least} one of them must be wrong. Since
$N$-body calculation can suffer many numerical problems due to limited
resolution and particle noise, one might think $N$-body result is
probably wrong. However, additional tests with different number of
particles and different sizes of timestep showed very good agreement
(Hashimoto \etal, 2002). Therefore it seems our $N$-body result is
sound. In addition, as we stated in the previous section, our $N$-body
result is in good agreement with BLLS and JB. Though it is not
impossible, it is certainly unlikely that all of these three works are
wrong.

So let us now consider the possibility that the standard dynamical
friction formula is wrong.

The standard dynamical friction formula is obtained under the
assumption that the massive object moves straight in a uniform and
isotropic distribution of field particles. Field particles are also
assumed to be moving straight, and any interaction between field
particles is ignored. Clearly, the satellite does not move straight,
but circle around the center of the parent galaxy. The distribution of 
field stars within the parent galaxy is far from uniform, and field
stars also circle around in the parent galaxy. Thus, it is not really
surprising that the naive use of the dynamical friction formula gives
rather bad result.

One obvious way to improve the accuracy of the dynamical friction
formula is to calculate the linear response of the global distribution 
function of the parent galaxy to the presence and the orbit of the
satellite (Weinberg, 1995). This approach would certainly
give accurate and reliable result which agrees well with $N$-body
result (Hernquist and Weinberg 1989).  However, since the global
response depends on the distribution function itself, the result
cannot be expressed in a compact and form. So here we consider the
possibility to improve the standard formula.

As we noted above, there are at least two problems with the standard
formula. First, it assumes that both the satellite and field stars
move straight. Second, it assumes that the density of the field star
is the same everywhere.

The first assumption is clearly wrong, but its effect is difficult to
estimate. Let us consider the effect of the second assumption, which
is much easier to evaluate. In previous works, the outer cutoff radius 
of the Coulomb logarithm is taken to be the scale length of the halo,
while the representative density of the field stars is taken to be the 
local density around the satellite. This would clearly cause an
overestimate of the Coulomb integral, for the case of the singular
isothermal sphere (or the King model we used), since the stellar
density drops off as fast as $1/r^2$. This means the logarithmic
divergence of the  Coulomb integral does not actually occur if we
takes into account the effect of the density gradient.

To correctly take into account the effect of the density gradient is a
tricky problem.  We cannot really use the straight line approximation
for encounters with impact parameter comparable to or larger than
$R_s$, the distance to the center of the galaxy.  On the other hand,
it might not be too bad an assumption just to ignore any encounter with
impact parameters comparable to or larger than $R_s$.  The density
drops off rapidly and realistic effect is unlikely to enhance the
effect of the encounter (except for the small fraction of the orbits
in resonance with the orbit of satellite).

Thus, it might be more sensible to use $R_s$ as the outer cutoff
radius for the Coulomb logarithm, that to use the traditional
$R_{halo}$. In fact, this use of $R_s$ is first proposed by a
pioneering work by Tremaine (1976) on the effect of the dynamical
friction to the orbit of LMC-SMC pair.

To use $\epsilon_s$ as the inner cutoff is okay as an
order-of-magnitude estimate, but can be improved by actually
integrating the effect of all encounters with small impact parameters
for Plummer potential, following the treatment by White (1976). For
Plummer model, the integration can be performed analytically and the
result is that inner cutoff radius is $r_{in}=1.6 \epsilon_s$.

\begin{figure}
\plotone{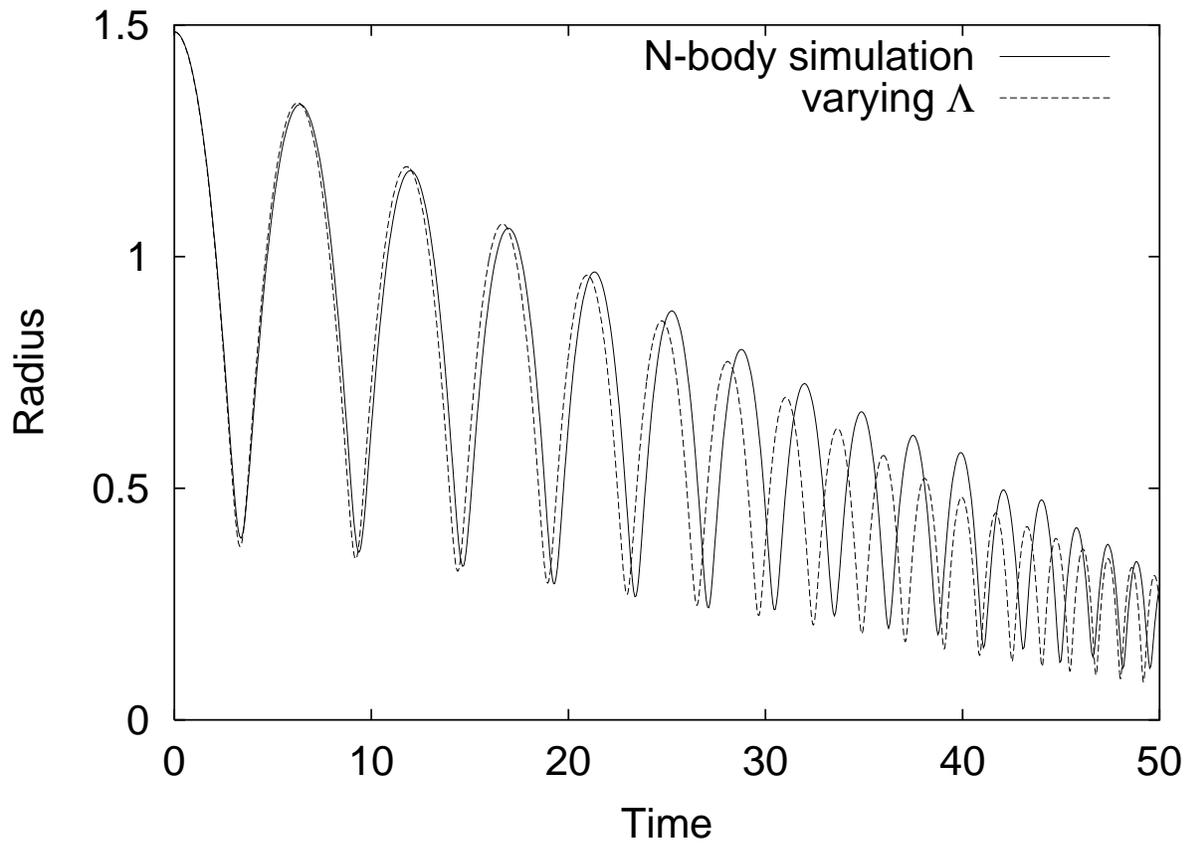}
\caption{Same as Figure 1, but for the variable $\Lambda$.}
\label{fig:2}
\end{figure}

Figure \ref{fig:2} shows the result of the $N$-body simulation
compared to that of our improved (both outer and inner cutoffs)
semi-analytic treatment.  Figure \ref{fig:2} is the same as Figure
\ref{fig:1} but for the
choice of the Coulomb logarithm
\begin{equation}
\ln \Lambda = \ln \left(\frac{R_{\mathrm s}}
{1.4 \epsilon_{\mathrm s}}\right).
\label{eqn:vlambda}
\end{equation}
During our semi-analytic integration using (\ref{eqn:vlambda}), when
the $R_{\mathrm s}$ becomes smaller than $1.4 \epsilon_{\mathrm s}$,
we simply put the dynamical friction term to be zero, since it is
clearly unphysical to apply dynamical ``acceleration''.

Figure \ref{fig:2} shows that the results of the $N$-body simulation
and semi-analytic treatment agree quite well.

The coefficient appeared in the denominator of the right-hand side of
equation (\ref{eqn:vlambda}) is chosen in order to fit the
curve of our semi-analytic model to that  of $N$-body simulations.
%
%
The Coulomb logarithm which we used to fit the result of direct
$N$-body simulation is in good agreement with the that proposed by
White (1976).  Strictly speaking, there is a slight difference between
the value of the coefficient in the denominator we used in Figure
\ref{fig:2} and that estimated by the method proposed in White (1976) :
the former is 1.4, while the latter is 1.6.  It is not serious. The
latter value is estimated using straight-line approximation so that
the value should be overestimated a little. (In other words, the
straight-line approximation tends to underestimate the effect of
dynamical friction.) We will discuss this problem elsewhere
(Hashimoto, \etal, 2002).

The agreement between the $N$-body result and semi-analytic treatment
is quite remarkable. 
Figure \ref{fig:2} shows that the discrepancy shown in Figure
\ref{fig:1} is caused by an
inadequate estimate of $\Lambda$. Other possible reasons, such as the
effect of the global response of the distribution function, might
still exists, but they are clearly not the prime reason of the
discrepancy between $N$-body and semi-analytic works which we
discussed in the introduction and section 3.

The improved agreement with the $N$-body result is explained as
follows. With $b_{max}=R_{cut}$, the semi-analytical treatment causes
strong circularization and faster orbital evolution. This implies that 
the the semi-analytical treatment overestimated the dynamical friction 
around the perigalacticon. Around the apogalacticon, the error might exist,
but relatively small compared to that at the perigalacticon. The use of
variable $b_{max}$ reduces the value of $\ln \Lambda$ both at
perigalacticon and apogalacticon, but by a much larger factor at the
perigalacticon simply because $R_s$ is smaller. Thus, effectively we
reduced the dynamical friction around the perigalacticon, which resulted
in the improvement in the agreement with the $N$-body result.

In hindsight, it looks too obvious that the traditional use of the
dynamical friction formula was inappropriate. Theoretically, it is
clearly not justifiable to assume that the stellar density is the same
up to the outer cutoff radius of the halo. From comparison between the
$N$-body result and those of semi-analytic treatment, it also is clear
that previous semi-analytic treatment overestimates deceleration due
to the dynamical friction around the perigalacticon.

To summarize our result, the orbital decay of satellites is slower
than ever estimated, the eccentricity of orbit of revolution of a
satellite around the host galaxy is almost constant. The reason why
previous estimates are wrong is that previous studies overestimated
the effect of dynamical friction at the perigalacticon.


\section{Summary and Applications}

We performed $N$-body simulations of satellite orbits.  We found that
the circularization of the orbit due to the dynamical friction is much 
slower than commonly believed.  This discrepancy was also reported by
BLLS, and we can see the same tendency from the numerical result
reported by JB. 

Previous studies of satellite orbits used the outer cutoff radius of
the dark halo as $b_{max}$.   We found that the effective $b_{max}$ 
should be of the order of $R_s$, the distance of the satellite from the 
center of the galaxy, which varies as the satellite orbits around the
galaxy. Our formula results in a greatly improved agreement with the
$N$-body result.

\subsection{the Large Magellanic Cloud}

\begin{figure}
\plotone{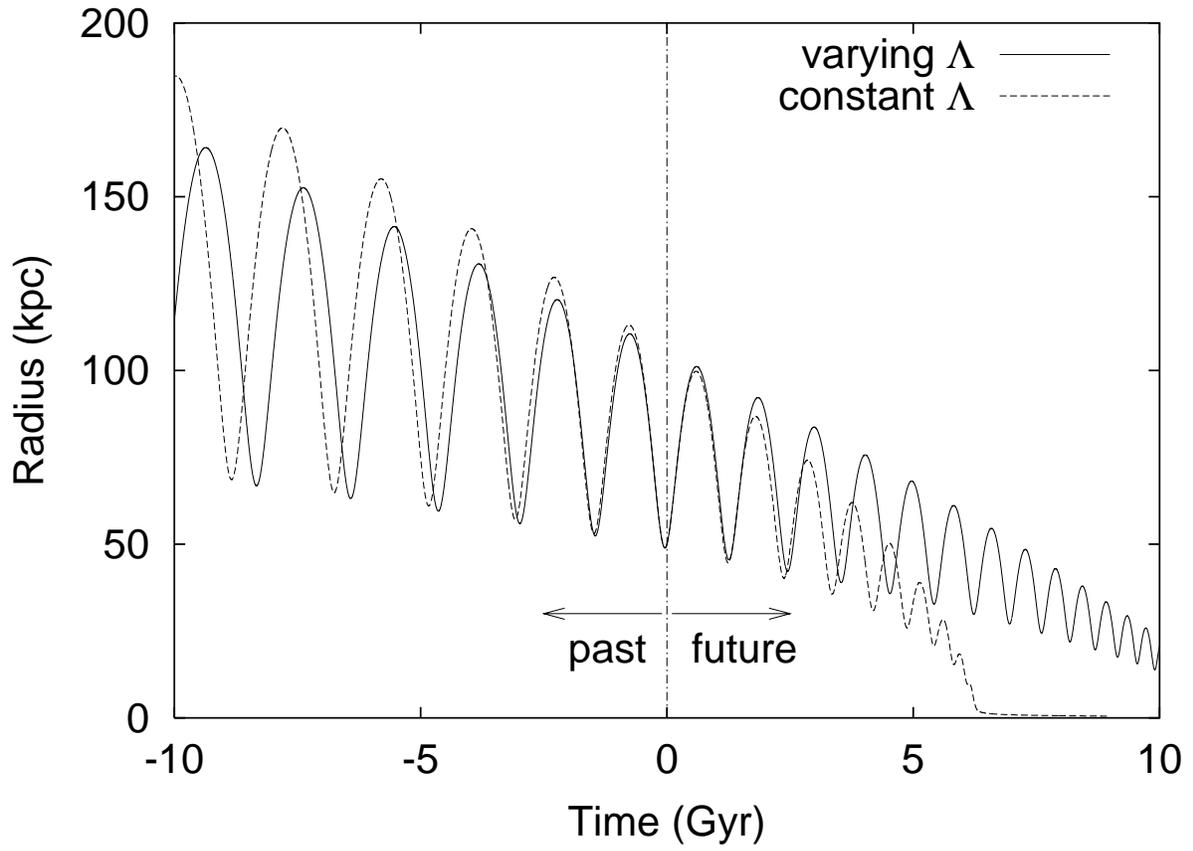}
\caption{Radial Evolution of LMC. From -10 G years to 10 G years.}
\label{fig:3}
\end{figure}

The Large Magellanic Cloud is the most famous satellite of Milky Way.
Its orbit has been investigated from both observation and
numerical simulations ({\it e.g.}, Toomre, 1970; Tremaine, 1976; Lin
and Lynden-Bell, 1977; Murai and Fujimoto, 1980).  The importance of
the effect of dynamical friction from the galactic halo on the orbit
evolution LMC is first emphasized by Tremaine (1976).

By using numerical simulation, \citet{MF} (hereafter MF) determined
the orbital elements and the present phase of the LMC.  They performed
a number of backward numerical integrations of the orbits of the LMC
and SMC from various initial conditions, and integrated orbits of test
particles in the LMC and SMC for each condition.  Comparing the result
of distribution of test particles and the observed Magellanic stream,
they chose the initial condition which gives the best fit.

In their numerical integration, they assumed a halo expressed by a
singular isothermal sphere, which is a simple flat-rotation halo.  In
their paper, it is not clear either what assumption or what exact
value is adopted for $\ln \Lambda$, since there is no discussion on
how they determined $\ln \Lambda$ though it appeared in their equation
(13).

In order to see the effect of changing $\ln \Lambda$, we integrated the
orbit of LMC both forward and backward in time, using both the
constant $\Lambda$ and variable $\Lambda$ ($b_{max} = R_s$). In this
study, we express LMC as a single Plummer-softened particle with mass
$2 \times 10^{10}\smass$ and softening length $5$ kpc.  The rotation velocity of the halo is
$250$ km/s, same as what is used by MF. We simulated the orbit of
the LMC only, since our purpose here is to demonstrate the effect of
$\Lambda$ and not the accurate determination of the orbits of the
Clouds.

The solid curve in Figure \ref{fig:3} corresponds to the orbit
obtained when the dynamical friction is calculated using equation
(\ref{eqn:vlambda}). The dashed curve Figure \ref{fig:3} correspond to
the orbit obtained using the formula (\ref{eqn:dff1}) and
(\ref{eqn:clambda}).  Note that the backward part of this dashed curve
is in very good agreement with the result of MF. This agreement
strongly
suggests that what MF used is indeed a constant $\Lambda$.

 
Figure \ref{fig:3} shows that real evolution of the orbit of LMC (with
variable $\Lambda$) is significantly smaller than what is obtained by
MF. 10 Gyrs ago, the ``true'' apogalacticon was only $160$ kpc, while
the solution by MF was $180$ kpc.

A more remarkable difference is in the future of the LMC. 
With the constant $\Lambda$. 
The LMC  will fall to the galactic center in only 7 Gyrs with constant 
$\Lambda$, while our result suggests that it will take more than 14 G
years for the  LMC to fall to the galactic center.

\subsection{Statistical Evolution of Faint Galaxies}

In semi-anaritic studies of galaxy formation, it has
been assumed that the orbits of satellite galaxies evolve through
dynamical friction following Chandrasekhar's formula with constant
$\Lambda$.  In this section, we discuss how our result might change
our understanding of the statistical evolution of the satellite
galaxies.

Our study shows that the time evolution of the eccentricity of
satellites is rather small. Thus, we may assume that the distribution
of eccentricities of satellite galaxies at present directly reflects
that at the formation epoch of the Galaxy. Therefore the distribution
of eccentricities of satellites galaxies can be an important clue to
the formation of the Galaxy.

The lifetime of the satellite is estimated using the dynamical
friction timescale with $\ln \Lambda$ taken to be $M_H/M_s$
\citep{Lac93,Kau94}.  This would cause a quite serious overestimate in
the dynamical friction timescale, since the factor one should use is
the ratio between the size of the halo and the size of the satellite.
If we assume $M\propto \sigma^4$, we have $R \propto M^{1/2}$. Thus,
there is at least a factor of two difference in the value of $\ln
\Lambda$.  Since there are too many other uncertainties in the
semi-analytic modeling of the galaxy number evolution, how serious
this difference is not clear. However, it certainly affects the
estimate of presently observed satellites rather strongly. A more
detailed study on this aspect is clearly necessary.

We thank Toshi Fukushige and Sadanori Okamura on stimulating
discussions. We also thank Rainer Spurzem, T, Tsuchiya, and Andrea
Just for helpful discussions. This work is supported by Grant-in-Aid
for Scientific Research B (13440058) of the Ministry of Eduaction,
Culture, Culture, Science and Technology, Japan.

\begin{deluxetable}{ccccccccccc}
\tabletypesize{\scriptsize}
\tablecaption{Model Parameters of $N$-body simulations \label{tbl-1}}
\tablewidth{0pt}
\tablehead{ 
\colhead{Galactic Halo} & 
\colhead{${}^aM_{{\rm gal}}$} & 
\colhead{${}^b\epsilon_{\rm halo}$}&
\colhead{${}^cM_{{\rm sat}}$} &
\colhead{${}^d\epsilon_{{\rm sat}}$} &
\colhead{Initial Position} &
\colhead{Initial Velocity} &
}
\startdata
King model ($\Psi_0=9$)& 1.0&0.0315&0.01&0.1&(1.5,0)&(0,0.326)\\
\enddata
\tablenotetext{a}{Total mass of the host galaxy} 
\tablenotetext{b}{Softening length of a halo particle}
\tablenotetext{c}{Satellite mass}
\tablenotetext{d}{Softening length of the satellite}


\end{deluxetable}

\begin{deluxetable}{crrrrrrrrrr}
\tabletypesize{\scriptsize}
\tablecaption{Model Parameters of LMC\label{tbl-2}}
\tablewidth{0pt}
\tablehead{
\colhead{ run } & 
\colhead{Galactic Halo} & 
\colhead{$M_{{\rm gal}}$} & 
\colhead{${}^aV_c$} & 
\colhead{$M_{{\rm sat}}$} & 
\colhead{$\epsilon_{{\rm sat}}$}  &
\colhead{Initial Position}&
\colhead{Initial Velocity} &
}
\startdata
LMC &Singular Isothermal Sphere& ${}^b7.3\times 10^{11}\smass$
	&$250$ km/s&$2\times 10^{10} \smass$
	&5 kpc&($50$kpc ,$0$kpc)&($50$km/s,$340$km/s)\\
\enddata

\tablenotetext{a}{Constant circular speed of this galactic model}
\tablenotetext{b}{Halo mass within $50$kpc from the galactic center}


\end{deluxetable}
\end{document}